\documentclass[a4paper,12pt]{article}
\usepackage{amssymb,textcomp,amsmath,graphicx,bm,subfig,comment,hyperref,float,color,cite}
\usepackage[flushleft]{threeparttable}
\usepackage{array,booktabs,makecell}
\usepackage[perpage]{footmisc}

\long\def\symbolfootnote[#1]#2{\begingroup
	\def\thefootnote{\fnsymbol{footnote}}
	\footnote[#1]{#2}\endgroup}

\begin{document}
\def\gsim{\:\raisebox{-0.5ex}{$\stackrel{\textstyle>}{\sim}$}\:}
\def\lsim{\:\raisebox{-0.5ex}{$\stackrel{\textstyle<}{\sim}$}\:}
\pagestyle{empty}
\vspace*{1.5cm}

\begin{center}
  {\Large \bf Impact of the Bounds on the Direct Search for Neutralino
    Dark Matter on Naturalness} \\
  \vspace{1cm}
  {\large Manuel Drees$^1$\symbolfootnote[0]{$^*$drees@th.physik.uni-bonn.de}$^*$ and Ghazaal Ghaffari$^{1,2}$\symbolfootnote[0]{$^\dagger$ghazalf@uni-bonn.de, ghafari\_ghazaal@physics.sharif.edu}$^\dagger$} \\
  \vspace*{6mm} {${}^1$ Bethe Center for Theoretical Physics and
    Physikalisches Institut, Universit\"at Bonn,\\Nussallee~12,
    D-53115 Bonn, Germany\\
    ${}^2$ Department of Physics, Sharif University of Technology}
\end{center}

\begin{abstract}
  In the Minimal Supersymmetric Extension of the Standard Model (MSSM)
  the higgsino mass parameter $\mu$ appears both in the masses of the
  Higgs bosons and in the neutralino mass matrix. Electroweak
  finetuning therefore prefers small values of $|\mu|$. On the other
  hand, bino--like neutralinos make a good dark matter candidate. We
  show that current direct search limits then impose a strong lower
  bound on $|\mu|$, in particular for $\mu > 0$ or if the masses of
  the heavy Higgs bosons of the MSSM are near their current limit from LHC
  searches. There is therefore some tension between finetuning and neutralino
  dark matter in the MSSM. We also provide simple analytical expressions
  which in most cases closely reproduce the numerical results.
\end{abstract}
\clearpage

\setcounter{page}{1}
\pagestyle{plain}
\tableofcontents
\section{Introduction}

Softly broken supersymmetry alleviates the electroweak hierarchy
problem \cite{Witten:1981nf,Sakai:1981gr}. In the limit of exact
supersymmetry there are no quadratically divergent radiative
corrections to the masses of Higgs bosons. Increasing lower bounds of
the masses of superparticles from searches at the LHC \cite{pdg} imply
sizable loop corrections to the masses of Higgs bosons in realistic
supersymmetric models. This leads to finetuning if soft breaking
masses are treated as uncorrelated \cite{Barbieri:1987fn,
  Ellis:1986yg}. However, this source of finetuning might be greatly
reduced in models with a small number of independent soft breaking
parameters, typically defined at a very large renormalization scale,
which introduce correlations between weak--scale parameters
\cite{Baer:2012cf}.

On the other hand, the minimal supersymmetric extension of the
standard model (MSSM) \cite{Haber:1984rc,Nilles:1983ge} also requires
a supersymmetric contributions to Higgs and higgsino masses; this mass
parameter is usually called $\mu$. Searches for charginos imply
\cite{pdg} $|\mu| \gsim 100$ GeV. Since $\mu$ enters the Higgs
potential already at tree--level, a value of $|\mu|$ much above $M_Z$
inevitably leads to finetuning. There is thus general agreement in the
discussion of finetuning issues in the MSSM that -- given the
experimental lower bound -- smaller values of $|\mu|$ are preferred,
with the necessary degree of finetuning increasing like $\mu^2$.

The same parameter $\mu$ also sets the mass scale for the higgsinos in
the MSSM;\footnote{This can be avoided only if one introduces additional
  non--holomorphic soft--breaking higgsino mass terms \cite{Ross:2017kjc};
  however, most supersymmetry breaking mechanisms do not generate such
  terms. We will therefore not consider this possibility here.} as already
noted, this is the origin of the lower bound on $|\mu|$. This establishes
a connection between finetuning and the phenomenology of the neutralinos
and charginos in the MSSM.

One of the attractive features of the MSSM with exact $R-$parity is
that it automatically contains a candidate particle to form the
cosmological dark matter whose existence can be inferred from a host
of observations, assuming only that Einsteinian (or indeed Newtonian)
gravity is applicable at the length scales of (clusters of) galaxies
\cite{Bertone:2004pz}. This candidate is the lightest neutralino
$\tilde \chi_1^0$ \cite{Jungman:1995df}, whose mass is bounded from
above by $|\mu|$. Given that naturalness arguments prefer a small
value of $|\mu|$, one might assume that the most natural dark matter
candidate in the MSSM is a light higgsino--like neutralino. However,
in minimal cosmology a higgsino--like neutralino has the correct relic
density only for $|\mu| \simeq 1.2$ TeV \cite{Edsjo:1997bg}, which
would lead to permille--level electroweak finetuning. A wino--like LSP
would have to be even heavier. One can appeal to non--standard
cosmologies, e.g.  including non--thermal production mechanisms, in
order to give a lighter higgsino--light neutralino the correct relic
density; however, such scenarios are already quite strongly constrained
by indirect dark matter searches \cite{Baer:2018hpb}.

In this article we therefore assume that the bino mass parameter
$M_1 < |\mu|$, so that the LSP eigenstate is dominated by the bino
component; $M_1$ can be taken positive without loss of
generality. Also in this case finetuning would prefer $|\mu|$ to be
not far above $M_1$. On the other hand, if $|\mu| \simeq M_1$ the LSP
has sizable higgsino and bino components, and hence generically
sizable couplings to the neutral Higgs bosons of the MSSM. Such mixed
neutralinos tend to have rather large scattering cross sections on
matter, in potential conflict with strong lower bounds on this
quantity from direct dark matter searches \cite{pdg}. This is true in
particular for the so--called ``well tempered neutralino''
\cite{ArkaniHamed:2006mb}, a bino--higgsino mixture with the correct
relic density in minimal cosmology.

In this article we explore this connection quantitatively, by deriving a
lower bound on the difference $|\mu| - M_1$ from the upper bound on
the neutralino--nucleon scattering cross section found by the Xenon
collaboration \cite{Aprile:2018dbl}. We do this both numerically, and
using a simple approximation for the bino--like neutralino eigenstate
which is very accurate in the relevant region of parameter space. The
resulting lower bound on $|\mu|$ is much stronger than the trivial
constraint $|\mu| > M_1$ which follows from the requirement of a
bino--like LSP; this is true in particular if $M_1$ and $\mu$ have the
same sign. An upper bound on $|\mu|$ from finetuning considerations
therefore leads to a considerably stronger upper bound on $M_1$ from
direct dark matter searches.

A bino--like neutralino will often have too large a relic density in
minimal cosmology. This can be cured either by assuming non--standard
cosmology (e.g. a period of late entropy production)
\cite{Kamionkowski:1990ty, Gelmini:2006pw}, or -- for not too small
$M_1$ -- by arranging for co--annihilation with a charged
superparticle, e.g. a $\tilde \tau$ slepton \cite{Ellis:1999mm}, which
can still have escaped detection by collider experiments if it is
close in mass to the lightest neutralino. Neither of these
modifications changes the cross section for neutralino--proton
scattering significantly. By not imposing any relic density constraint
our result thus becomes less model--dependent. This, as well as the
use of the more recent, and considerably stronger, Xenon--1T
constraint and the approximate analytical derivation of the constraint
on $|\mu|$, distinguishes our analysis from that of
ref.\cite{Profumo:2017ntc}.

The rest of this article is organized as follows. In the following section
we review neutralino mixing, both exact and using a simple approximation.
We also give the relevant expressions for the neutralino--nucleon scattering
cross section. In section~3 we present the resulting lower bound on the
difference $|\mu| - M_1$ as a function of $M_1$, before concluding in
section~4.

\section{Formalism}

In this section we briefly review the neutralino masses and mixings
in the MSSM, as well as the spin--independent contribution to
neutralino--nucleon scattering.

\subsection{The Neutralinos in the MSSM}

The neutralinos are mixtures of the two neutral gauginos (the bino
$\tilde B$ and the neutral wino $\widetilde W_3$) and two neutral
higgsinos $\tilde h_d^0, \, \tilde h_u^0$ associated with the two
$SU(2)$ Higgs doublets required in the MSSM.  The resulting mass
matrix in the
$\tilde B,\, \widetilde W_3, \, \tilde h_d^0,\, \tilde h_u^0$ basis is
given by \cite{Nilles:1983ge}:
\begin{equation} \label{eq:matrix}
{\cal M}_0 = \mbox{$ \left( \begin{array}{cccc}
M_1 & 0 & - M_Z \cos \! \beta \sin \! \theta_W & M_Z \sin \! \beta \sin \!
\theta_W \\
0 & M_2 &   M_Z \cos \! \beta \cos \! \theta_W & -M_Z \sin \! \beta \cos \!
\theta_W \\
- M_Z \cos \! \beta \sin \! \theta_W & M_Z \cos \! \beta \cos \! \theta_W & 
0 & -\mu  \\
M_Z \sin \! \beta \sin \! \theta_W & -M_Z \sin \! \beta \cos \! \theta_W &
-\mu  & 0
\end{array} \right)\,. $} 
\end{equation}
Here $M_1$ and $M_2$ are soft breaking masses for the bino and wino,
respectively, $\mu$ is the higgsino mass parameter, $\theta_W$ is the
weak mixing angle, $M_Z \simeq 91$ GeV is the mass of the $Z$ boson,
and $\tan\beta = \langle H_u^0 \rangle / \langle H_d^0 \rangle$ is the
ratio of the vacuum expectation values (VEVs) of the two neutral Higgs
fields. The mass matrix is diagonalized by the $4 \times 4$ matrix
$N$, such that the $i-$th neutralino eigenstate is given by
\begin{equation} \label{eq:es}
  \tilde \chi_i^0 = N_{i1} \tilde B + N_{i2} \widetilde W_3 +
  N_{i3} \tilde h_d^0 + N_{14} \tilde h_u^0\,.
\end{equation}

Here we are interested in scenarios where the lightest neutralino
$\tilde \chi_1^0$, which is our dark matter candidate, is dominated by
the bino component. This requires $|M_1| < |M_2|, \, |\mu|$. We will
assume that these mass parameters are real; nontrivial complex phases
would contribute to CP violation, which is strongly constrained by
upper bounds on the electric dipole moments of the electron and
neutron \cite{Arbey:2019pdb}. Without loss of generality we take $M_1$
to be positive, but allow both signs for $\mu$. $|M_2|$ is constrained
significantly by searches for charginos and neutralinos at the LHC
\cite{pdg}; as long as $|M_2| > M_1$ the sign of $M_2$ is essentially
irrelevant for our analysis.

Evidently the mixing between the gaugino and higgsino states is
controlled by the mass of the $Z$ boson. If the difference between the
gaugino masses and $|\mu|$ is larger than $M_Z$, all mixing angles
will therefore be quite small, allowing for an approximate
perturbative diagonalization of the mass matrix (\ref{eq:matrix}). In
particular, the components of a bino--like $\tilde \chi_1^0$ can then be
approximated by \cite{Drees:1992am, Choi:2001ww}
\begin{equation} \label{eq:es-approx}
\begin{split}
  N_{12} &\simeq -M_Z^2 \cos{\theta_W} \sin{\theta_W}
  \frac {M_1 +\mu \sin{2\beta}} {(M_1-M_2) (M_1^2-\mu^2)} \,; \\
  N_{13} &\simeq -M_Z \sin{\theta_W}
  \frac {M_1 \cos{\beta} + \mu \sin{\beta}} {(M_1^2-\mu^2)} \,; \\
  N_{14} &\simeq M_Z \sin{\theta_W}
  \frac{M_1 \sin{\beta} +\mu \cos{\beta}} {(M_1^2-\mu^2)} \,; \\
 N_{11} &= \sqrt{1-N_{12}^2-N_{13}^2-N_{14}^2} \,.
\end{split}
\end{equation}

\begin{figure}[ht]
\centering
\includegraphics[width=0.8\textwidth]{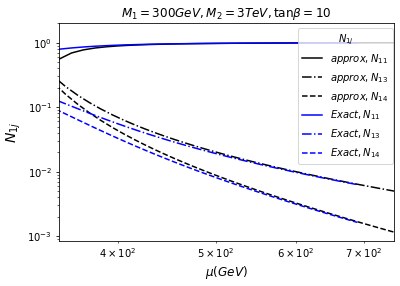}
\caption {The bino and higgsino components of the lightest
  neutralino eigenstate as a function of $\mu$, for fixed $M_1 = 300$
  GeV, $M_2 = 3$ TeV and $\tan\beta=10$. The simple approximation of
  eqs.(\ref{eq:es-approx}), shown in black, describes the exact (blue)
  results very well once $\mu - M_1 \geq 2 M_Z$.}
 \label{fig:Nij}
\end{figure}

In Fig.~\ref{fig:Nij} we compare these approximate expression with
exact results. Evidently the approximation works very well for
$\mu - M_1 \geq 2 M_Z$ or so. Since we took a very large value of
$M_2$ the wino component essentially vanishes in this example;
however, the first eq.(\ref{eq:es-approx}) shows that it only appears
at second order in $M_Z$, and is therefore always much smaller than
the higgsino components in the region of interest. In this figure we
have chosen $\mu$ to be positive. The second and third
eq.(\ref{eq:es-approx}) shows that this increases the higgsino
components. As a result, for $\mu < 0$ the approximation
(\ref{eq:es-approx}) becomes very accurate already for
$|\mu| - M_1 \geq 1.5 M_Z$.

\subsection{Neutralino--Nucleon Scattering in the MSSM}
  
In the limit of vanishing neutralino velocity only two kinds of
interactions contribute to neutralino--nucleon scattering. One of them
depends on the spin of the target nucleus; this contribution is
usually sub--dominant for heavy target nuclei like Xenon or
Germanium, which currently yield the tightest constraints for
neutralino masses above 10 GeV or so. The spin independent contributions
dominate because their contribution to the scattering cross section of
heavy nuclei scales like the square of the nucleon number. They originate
from the effective Lagrangian \cite{Jungman:1995df}
\begin{equation} \label{eq:Leff}
L_{\rm SI}^{\rm eff} = f_q \overline{\tilde\chi_1^0} \tilde\chi_1^0 \bar{q}q
\end{equation}
which describes the interactions of neutralinos with quarks. Here we
have limited ourselves to the leading, dimension--6, operator; strong
lower bounds on squark masses \cite{pdg} imply that dimension--8
operators due to squark exchange \cite{Drees:1993bu} can safely be
ignored.

Squark exchange also contributes at dimension $6$. However, this
contribution is proportional to the mass of the quark
\cite{Jungman:1995df}.  That is of course also true for Higgs
exchange contributions. However, at least the lighter neutral Higgs
boson, whose mass we now know to be $125$ GeV, lies about an order of
magnitude below the current lower bound on first generation squark
masses. Squark exchange diagrams therefore contribute only at the
$1\%$ level at best, which is well below the uncertainty of the Higgs
exchange contribution. We therefore ignore them in our analysis.

However, we do allow for the contribution of the heavier neutral Higgs
boson. The effective neutralino--quark coupling $f_q$ is thus given
by
\begin{equation} \label{eq:fq}
  f_q = \sum_{\phi=h,H} m_q \frac{g_{\phi \bar \chi \chi} g_{\phi \bar qq}}{m_{\phi}^2}\,.
\end{equation}
Note that we factored the quark mass out of the Higgs couplings to quarks,
making the latter independent of the quark mass; the couplings to
up-- and down--type quarks still differ, however. They are given by
\cite{Gunion:1984yn}:
\begin{equation} \label{eq:yuk}
\begin{split}
g_{h \bar uu} &= \frac{-g \cos\alpha}{2M_W \sin\beta} \simeq\frac{-g}{2M_W}\,;\\
g_{h\bar dd} &= \frac{g \sin\alpha}{2M_W \cos\beta}\simeq\frac{-g}{2M_W}\,;\\
g_{H\bar uu} &= \frac{-g \sin\alpha}{2M_W \sin\beta}
\simeq\frac{g}{2M_W\tan\beta}\,;\\
g_{H \bar dd} &= \frac{-g \cos\alpha}{2M_W \cos\beta}
\simeq \frac{-g\tan\beta}{2M_W}\,.
\end{split}
\end{equation}
Here $\alpha$ is the mixing angle between the two neutral Higgs
bosons, $g$ is the $SU(2)$ gauge coupling and $M_W \simeq 80$ GeV is
the mass of the $W$ boson. The couplings of the 125 GeV Higgs boson
are known to be quite close to those of the SM Higgs boson
\cite{pdg}. Moreover, none of the heavier Higgs bosons of the MSSM
have yet been found. Both observations can easily be satisfied in the
so--called decoupling limit, where the mass of the neutral CP--odd
Higgs boson satisfies $m_A^2 \gg M_Z^2$. In that limit the other heavy
MSSM Higgs bosons also have masses very close to $m_A$, and the mixing
angle $\alpha$ satisfies
$\cos\alpha \simeq \sin\beta,\ \sin\alpha \simeq -\cos\beta$. This
leads to the simplifications in the Higgs couplings to quarks given
after the $\simeq$ signs in eqs.(\ref{eq:yuk}). In particular, the
couplings of the lighter Higgs boson $h$ then approach those of the SM
Higgs, in agreement with observation. The couplings of the heavier
neutral Higgs boson $H$ to up--type quarks is suppressed by
$1/\tan\beta$, while its couplings to down--type quarks are enhanced
by $\tan\beta$.

The Higgs bosons couple to one gaugino and one higgsino current state.
As a result, their couplings to neutralino current eigenstates are
proportional to the product of gaugino and higgsino components
\cite{Gunion:1984yn}:
\begin{equation} \label{eq:ghcc}
\begin{split}
g_{h\bar\chi \chi} =& \frac{1}{2}(g N_{12} - g'N_{11})
  (N_{13}\sin\alpha + N_{14} \cos {\alpha}) \,; \\
  g_{H\bar\chi \chi} =& \frac{1}{2}(g N_{12} - g'N_{11})
  (N_{14} \sin\alpha - N_{13}\cos{\alpha}) \,,
\end{split}
\end{equation}
where $g'$ is the $U(1)_Y$ coupling.

The total spin--independent neutralino--proton scattering cross
section can be written as \cite{Jungman:1995df}
\begin{equation} \label{eq:cs}
\sigma_{SI}^{\chi p} = \frac{4\mu_{\chi p}^2} {\pi} |G_s^p|^2 \,,
\end{equation}
where
$\mu_{\chi p} = m_p m_{\tilde \chi_1^0}/ (m_p + m_{\tilde\chi_1^0})$
is the reduced mass of the neutralino--proton system, and the
effective neutralino--proton coupling is given by
\begin{equation} \label{eq:G}
G_s^p = -\sum_{q=u,d,..} \langle p| m_q\bar{q} q| p \rangle 
	\sum_{\phi=h,H} \frac{g_{\phi\bar\chi \chi} g_{\phi \bar qq}}{m_\phi^2} \,.
\end{equation}
For the light $u,d,s$ quarks, the hadronic matrix elements have to be
computed using non--perturbative methods. Once these are known, the
contribution from heavy $c,b,t$ quarks can be computed perturbatively
through a triangle diagram coupling to two gluons
\cite{Shifman:1978zn}. One usually parameterizes
$\langle N| m_q\bar{q} q| N \rangle = f_{T_q} m_p$. We use the
numerical values from DarkSUSY \cite{Bringmann:2018lay}:
\begin{equation} \label{eq:fT}
  f_{TU} \equiv \sum_{q = u,c,t} f_{Tq} = 0.14\,; \ \
  f_{TD} \equiv \sum_{q = d,s,b} f_{Tq} = 0.23\,.
\end{equation}
We note that these numbers are somewhat uncertain, but our values are
rather conservative \cite{Ellis:2018dmb}.

Putting everything together, using the approximate expressions
(\ref{eq:es-approx}) for the lightest neutralino eigenstate and
assuming the decoupling limit, we find:
\begin{equation} \label{eq:G-approx}
\begin{split}
\left. G_S^p \right|_h &\simeq -A \ m_p \left( f_{TU} + f_{TD} \right)
    \left( \frac{M_1+\mu \sin{2\beta}} {m_h^2 (\mu^2-M_1^2)} \right)\,; \\
\left. G_S^p \right|_H &\simeq -A \ m_p \left( \frac{ f_{TU}} {\tan\beta}
      - f_{TD} \tan\beta\right) \times \left(
      \frac{\mu \cos{2\beta}}{m_H^2(\mu^2-M_1^2)} \right)\,.
\end{split}
\end{equation}
Here we have introduced the constant
\begin{equation} \label{eq:A}
  A = \frac{g g' M_Z \sin\theta_W}{4 M_W} = 0.032\,.
\end{equation}
It should be noted that $\tan\beta > 1$ implies $\cos(2\beta) < 0$.
Eqs.(\ref{eq:fT}) imply that the term $\propto f_{TD}$ always
dominates $H$ exchange. Hence $h$ and $H$ exchange contribute with the
same sign if $\mu > 0$ or $\mu \sin2\beta < -M_1$; they interfere
destructively for $0 > \mu > -M_1/\sin2\beta$.

\section{Results}

We are now ready to present our numerical results. We wish to determine
the lower bound on $|\mu|$ that follows from the non--observation of
neutralinos, which we assume to form all of (galactic) dark matter.
We will not be concerned with very light neutralinos, where current bounds
from direct dark matter search are still quite poor \cite{pdg}. For
masses above $20$ GeV the most stringent current bound comes from the
Xenon--1T experiment \cite{Aprile:2018dbl}. In this range the bound
is well parameterized by
\begin{equation} \label{sig-max}
\begin{split}
\sigma^{\rm max}(m_{\tilde \chi_1^0})_{\rm XENON} &=
\left( \frac {m_{\tilde\chi_1^0}} {10 \ {\rm GeV}}
  + \frac {2.7 \cdot 10^4 \ {\rm GeV}^3 }{m_{\tilde \chi_1^0}^3} \right)
\cdot 10^{-47} \ {\rm cm}^2 \\
&= \left( \frac {m_{\tilde\chi_1^0}} {3.9 \ {\rm GeV}}
  + \frac {7 \cdot 10^4 \ {\rm GeV}^3 }{m_{\tilde \chi_1^0}^3} \right)
\cdot 10^{-20} \ {\rm GeV}^{-2}\,.
\end{split}
\end{equation}

\begin{figure}[ht]
\centering
\includegraphics[width=0.8\textwidth]{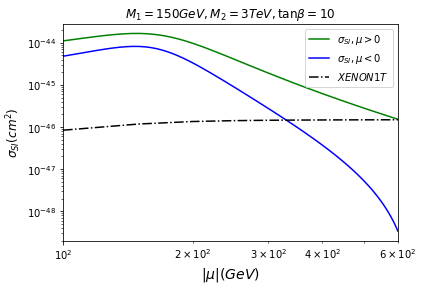}
\caption{The predicted neutralino--proton scattering cross section
  for $M_1 = 150$ GeV, $M_2 = 3$ TeV, $m_A = 1.8$ TeV and $\tan\beta = 10$
  as function of $|\mu|$, for positive (green) and negative (blue) $\mu$. The
bound from the Xenon--1T collaboration is shown as black dot--dashed line.}
\label{fig:fig3}
\end{figure}

Fig.~3 shows that this bound constrains the MSSM parameter space quite
severely, if we assume that the lightest neutralino $\tilde \chi_1^0$
forms all of dark matter. Here we have chosen $M_1 = 150$ GeV,
$m_A = 1.8$ TeV, $M_2 = 3$ TeV and $\tan\beta = 10$; the exact value
of $M_2$ is basically irrelevant as long as it is significantly larger
than $M_1$. As expected the predicted cross section is largest for
$|\mu| \simeq M_1$, which leads to strong bino--higgsino
mixing. However, the Xenon--1T bound requires $|\mu|$ well above
$M_1$, i.e. the lightest neutralino has to be bino--like; note that
values of $|\mu|$ below 100 GeV have not been considered here since
they are excluded by chargino searches at LEP for the given (large)
value of $M_2$. In the allowed range of $|\mu|$ the approximate
diagonalization of eqs.(\ref{eq:es-approx}) works quite
well. Eq.(\ref{eq:G-approx}) then explains why the lower bound on $|\mu|$
is considerably weaker for $\mu < 0$: evidently the two terms in the
numerator of $\left. G_S^p \right|_h$ tend to cancel (add up) for
positive (negative) $\mu$. As a result, for $\mu < 0$ the contribution
from $H$ exchange, which contributes with opposite sign than the
dominant $h$ exchange term, is relatively more important and further
reduces the cross section. In fact, the $h$ exchange contribution
vanishes at $\mu = -M_1/\sin(2\beta) \simeq 760$ GeV. Due to $H$
exchange the actual zero of the cross section \cite{Drees:1993bu} --
the so--called ``blind spot'' \cite{cheung2013} -- already occurs at
$\mu \simeq -660$ GeV. In contrast, for $\mu > 0$ the (small) $H$
exchange contribution slightly strengthens the lower bound on $\mu$,
which saturates at $\sim 590$ GeV for $m_H \rightarrow \infty$. This
value is already uncomfortably large in view of finetuning
considerations.

\begin{figure}[ht]
\centering
\includegraphics[width=0.8\textwidth]{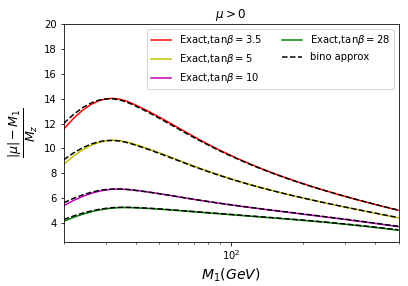}
\caption{The lower bound on $|\mu| - M_1$, in units of $M_Z$, that
  follows from the upper bound on the neutralino--proton scattering
  cross section derived by the Xenon--1T collaboration, for $\mu >
  0$. We have again chosen $M_2 = 3$ TeV and $m_A = 1.8$ TeV. The
  solid lines show numerical results obtained using DarkSUSY for
  different values of $\tan\beta$, while the dashed curves are based
  on the approximate analytical diagonalization of the neutralino mass
  matrix, see eq.(\ref{eq:mubound}).}
\label{fig:fig4}
\end{figure}

This conclusion is reinforced by Fig.~\ref{fig:fig4}, which shows the
lower bound on $\mu - M_1$ in units of $M_Z$ as a function of $M_1$
for four different values of $\tan\beta$; the values of $M_2$ and
$m_A$ are as in Fig.~\ref{fig:fig3}. The solid colored lines have been
derived numerically using DarkSUSY, whereas the black dashed lines are
based on the approximate diagonalization of the neutralino mass
matrix. Recall that we ignore squark exchange, so that only $h$ and $H$
exchange contribute to the spin--independent scattering cross section.
Using eqs.(\ref{eq:G-approx}) the extremal values of $\mu$ that
saturate the experimental upper bound on the cross section can be
computed analytically. To this end we introduce the quantities
\begin{equation} \label{eq:pars}
  \begin{split}
    \kappa &= \frac {\sqrt{\pi \sigma^{\rm max}}} {2 A m_p^2}\,; \\
    c_\mu &= \frac { (f_{TD} + f_{TU}) \sin 2\beta} {m_h^2}
    - \frac{f_{TD} \tan\beta - f_{TU} \cot\beta } {m_H^2}\,; \\
    c_1 &= \frac { f_{TD} + f_{TU}} {m_h^2}\,.
  \end{split}
\end{equation}
The quantity $A$ has been defined in eq.(\ref{eq:A}), and the
normalized hadronic matrix elements $f_{TD}$ and $f_{TU}$ in
eqs.(\ref{eq:fT}). As noted above, the contribution $\propto f_{TU}$
to the $H$ exchange contribution is essentially negligible. $c_\mu$
collects terms in the Higgs exchange amplitude that are proportional
to $\mu$; only $h$ exchange contributes to $c_1$, which gets
multiplied with $M_1$ in this amplitude. The extremal values of $\mu$
are then given by
\begin{equation} \label{eq:mubound}
  \mu_\pm = \frac {c_\mu} {2 \kappa} \pm \sqrt { \frac {c_\mu^2} {4 \kappa^2}
    + M_1^2 + \frac{c_1 M_1}{\kappa} }\,.
\end{equation}
The dashed lines in Fig.~\ref{fig:fig4} correspond to the positive
solution $\mu_+$.

We see that this approximation describes the numerical results very
well even for the smallest value of $M_1$ we consider. In particular,
the rather strong dependence on $\tan\beta$ originates from the
$\sin2\beta$ factor in the $h$ exchange contribution to $c_\mu$, see
the second eq.(\ref{eq:pars}); $H$ exchange is always subdominant for
our choice $m_H = 1.8$ TeV. The $\tan\beta$ dependence becomes
somewhat weaker for larger values of $M_1$, where the $c_1$ term
becomes more important which is independent of $\tan\beta$.

Evidently the Xenon--1T bound is quite constraining for $\mu > 0$. For
example, if we interpret electroweak finetuning considerations as
requiring $|\mu| < 500$ GeV, one finds $\tan\beta > 10$ for
$M_1 \geq 20$ GeV, and $M_1 < 115 \ (165)$ GeV for
$\tan\beta = 20 \ (50)$.

\begin{figure}[ht]
\centering
\includegraphics[width=0.8\textwidth]{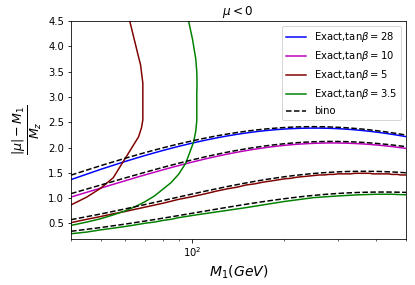}
\caption{The lower bound on $|\mu| - M_1$, in units of $M_Z$, that
  follows from the upper bound on the neutralino--proton scattering
  cross section derived by the Xenon--1T collaboration, for $\mu < 0$;
  the regions to the left of the upper red and green curves are also
  excluded. Parameter values and notation are as in
  Fig.~\ref{fig:fig4}.}
\label{fig:fig5}
\end{figure}

In eq.(\ref{eq:mubound}) we have assumed that
$c_\mu \mu + c_1 M_1 > 0$, which is always true for $\mu > 0$. It
remains true for values of $|\mu|$ below the ``blind spot''; if $H$
exchange is negligible this corresponds to $|\mu| \sin2\beta <
M_1$. The negative solution in eq.(\ref{eq:mubound}) then gives the
value of $\mu$ where the cross section decreases below the lower bound
when coming from $\mu = 0$. Fig.~\ref{fig:fig5} shows that this again
describes the exact numerically derived bound quite well: although the
bound on $|\mu| - M_1$ is considerably weaker than for positive $\mu$,
we saw in eqs.(\ref{eq:es-approx}) that there are cancellations in the
small entries of the $\tilde \chi_1^0$ eigenstate if $\mu < 0$; since
corrections to this approximation are of order of the squares of these
small entries, for given $|\mu|$ the approximation works better for
$\mu < 0$. In this figure we only show results for $M_1 \geq 40$ GeV,
since for smaller values of $M_1$ the bound on $|\mu|$ often is below
chargino search limit of about $100$ GeV. Note that now the lower
bound on $|\mu|$ {\em in}creases with increasing $\tan\beta$. This is
because the ``blind spot'' $\mu = - M_1/\sin2\beta$ moves to larger
values of $|\mu|$ for larger $\tan\beta$. In the limit
$\tan\beta \rightarrow \infty$ the bound on $|\mu|$ becomes
independent of the sign of $\mu$.

Beyond the blind spot the sign of $c_\mu \mu + c_1 M_1$ flips. This
region of parameter space can still be described by
eq.(\ref{eq:mubound}), by simply setting $\kappa \rightarrow - \kappa$
everywhere. If $H$ exchange is negligible, this can easily make the
argument of the root in eq.(\ref{eq:mubound}) negative, signaling that
no solution exists. In this case the cross section remains below the
experimental bound for all values of $\mu$ below the $\mu_-$ solution
in the original eq.(\ref{eq:mubound}). For the parameters of
Fig.~\ref{fig:fig5} we find that this is true for $\tan\beta > 8$. For
smaller values of $\tan\beta$ a sizable region of parameter space (to
the left and below the red and green solid lines) beyond the blind
spot is again excluded.\footnote{These curves are reproduced very
  accurately by the approximate diagonalization of the neutralino mass
  matrix; we do not show these results in order not to clutter up the
  figure too much.}

So far we have assumed that the heavy Higgs boson is very heavy, so
that its contribution to neutralino--proton scattering is
subdominant. In fact, there are several constraints on the masses of
the heavy Higgs bosons in the MSSM, which can be characterized by the
mass of the CP--odd Higgs boson, $m_A$. The most robust bounds come
from direct searches for the heavy neutral Higgs bosons; the most
sensitive ones exploit their decay into $\tau^+\tau^-$ pairs. In
particular, a recent ATLAS analysis \cite{Aad:2020zxo} is sensitive to
$m_A$ up to about 2 TeV, for very large $\tan\beta$. For
$\tau^+\tau^-$ invariant masses around $400$ GeV there seems to be
some excess of events. While not statistically significant, it leads
to a bound which is somewhat worse than the expected
sensitivity.\footnote{CMS did not yet publish the corresponding
  analysis for the full Run--2 data set.}  We therefore chose a
parameter set just on the exclusion line, $m_A = 400$ GeV and
$\tan\beta = 8$, in order to illustrate the maximal possible effect
from heavy Higgs exchange. It should be noted that this choice leads
to a sizable contribution from charged Higgs boson loops to radiative
$b \rightarrow s \gamma$ decays \cite{Bertolini:1990if,
  Borzumati:1994kk}; however, these can be compensated by postulating
some amount of squark flavor mixing \cite{Okumura:2003hy}.

\begin{figure}[ht]
\centering
\includegraphics[width=0.475\textwidth]{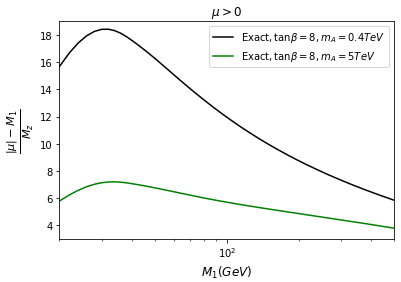}
\includegraphics[width=0.475\textwidth]{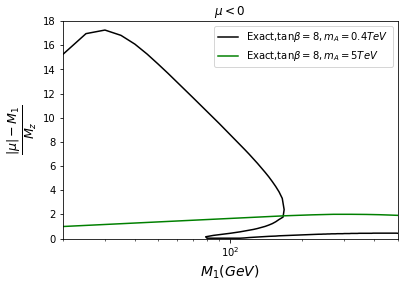}
\caption{The lower bound on $|\mu| - M_1$, in units of $M_Z$, that
  follows from the upper bound on the neutralino--proton scattering
  cross section derived by the Xenon--1T collaboration, for $\mu > 0$
  (left) and $\mu < 0$ (right). We have chosen $M_2 = 3$ TeV and
  $\tan\beta = 8$; the black (green) lines are for $m_A = 0.4 \ (5)$
  TeV.  In the right frame the enclosed region as well as the region
  above the first line are excluded.}
\label{fig:fig6}
\end{figure}

The bound on $|\mu|$ that results from the Xenon--1T constraint for
this choice of parameters is shown by the black lines in
Figs.~\ref{fig:fig6}; for comparison we also show results for
negligible $H$ exchange (green curves). As noted earlier for $\mu > 0$
both Higgs bosons always contribute with equal sign, so maximizing the
$H$ exchange contribution greatly strengthens the lower bound on
$\mu$. The effect is especially strong at smaller $M_1$ since $H$
exchange only contributes to $c_\mu$, not to $c_1$. The resulting
lower bound on $\mu$ is always above 950 GeV, i.e. in this region of
parameter space a bino--like lightest neutralino would yield little
benefit regarding electroweak finetuning compared with the canonical
thermal higgsino--like neutralino with $\mu \simeq 1.2$ TeV. Of
course, the black line shows the {\em maximal} effect from $H$
exchange. For somewhat larger $m_A$, away from the ATLAS lower bound,
the bound on $\mu$ will fall somewhere between the black and green
lines.

In sharp contrast, for $\mu < 0$ the $H$ exchange contribution reduces
the lower bound on $|\mu|$ even further, by moving the blind spot to
smaller values of $|\mu|$. However, for $M_1 \leq 170$ GeV the cross
section beyond the blind spot again increases above the Xenon--1T
bound, leading to a second excluded region. In fact, the allowed
region around the blind spot is very narrow for $M_1 \leq 150$ GeV.
The right frame of Fig.~\ref{fig:fig6} therefore shows that for
$m_A = 400$ GeV values of $M_1$ below about 150 GeV require
significant finetuning, either to hone in on the blind spot, or in the
electroweak sector due to the large values of $|\mu|$ required by the
Xenon--1T constraint away from the blind spot.

\section{Summary and Conclusions}

In this paper we have shown that the upper bound on the
neutralino--proton cross section from the Xenon--1T experiment leads
to strong lower bounds on $|\mu|$ if the bino mass parameter $M_1$
exceeds 20 GeV. We have assumed that the lightest neutralino forms all
of dark matter, but did not require the correct thermal relic density
in minimal cosmology. This constraint causes tension with electroweak
finetuning arguments, since in the MSSM with holomorphic soft breaking
terms the higgsino mass parameter $\mu$ also contributes to the masses
of the Higgs bosons. The bound is usually significantly stronger for
$\mu > 0$; however, also for $\mu < 0$ we found very strong
constraints if the heavy Higgs bosons are close in mass to current
experimental constraints and $M_1 \leq 150$ GeV. This argument can be
turned around to derive an upper bound on $M_1$ from an upper bound on
$|\mu|$ from electroweak finetuning; the latter is, however, not easy
to quantify unambiguously \cite{Drees:1995hj}. Our analytical
expressions will help to easily update these constraints when future
direct dark matter searches are published.

The Xenon--1T bound becomes considerably weaker for neutralino masses
below 20 GeV, which we did not consider in this paper. While even very
small values of $M_1$ remain experimentally allowed as long as all
sfermions are sufficiently heavy \cite{Dreiner:2009ic}, they do not
appear particularly plausible given the ever strengthening lower
bounds on the masses of the other gauginos (electroweak winos as well
as gluinos), chiefly from searches at the LHC \cite{pdg}. In fact,
many models of supersymmetry breaking predict fixed ratios between
these masses \cite{Choi:2007ka}, leading to strong lower bounds on
$M_1$. Our analysis would then lead to even stronger lower bounds on
$|\mu|$.

\subsubsection*{Acknowledgments}
GG thanks the Deutsche Akademische Auslandsdienst (DAAD) for partial financial
support and the Bethe Center for Theoretical Physics (BCTP) for its
hospitality. We thank Hessamaddin Arfaei for discussions.


\begin{thebibliography}{99}

\bibitem{Witten:1981nf}
E.~Witten,
Nucl. Phys. B \textbf{188} (1981) 513 doi:10.1016/0550-3213(81)90006-7.

\bibitem{Sakai:1981gr}
N.~Sakai,
Z. Phys. C \textbf{11} (1981) 153, doi:10.1007/BF01573998.

\bibitem{pdg}
  P.A. Zyla et al. (Particle Data Group), Prog. Theor. Exp. Phys. 2020,
  083C01 (2020). 

\bibitem{Barbieri:1987fn}
R.~Barbieri and G.~F.~Giudice,
Nucl. Phys. B \textbf{306} (1988) 63, doi:10.1016/0550-3213(88)90171-X.

\bibitem{Ellis:1986yg}
J.~R.~Ellis, K.~Enqvist, D.~V.~Nanopoulos and F.~Zwirner,
Mod. Phys. Lett. A \textbf{1} (1986) 57, doi:10.1142/S0217732386000105.

\bibitem{Baer:2012cf}
H.~Baer, V.~Barger, P.~Huang, D.~Mickelson, A.~Mustafayev and X.~Tata,
Phys. Rev. D \textbf{87} (2013) 115028, doi:10.1103/PhysRevD.87.115028
[arXiv:1212.2655 [hep-ph]].

\bibitem{Haber:1984rc}
H.~E.~Haber and G.~L.~Kane,
Phys. Rept. \textbf{117} (1985) 263, doi:10.1016/0370-1573(85)90051-1.

\bibitem{Nilles:1983ge}
H.~P.~Nilles,
Phys. Rept. \textbf{110} (1984) 1, doi:10.1016/0370-1573(84)90008-5.

\bibitem{Ross:2017kjc}
G.~G.~Ross, K.~Schmidt-Hoberg and F.~Staub,
JHEP \textbf{03} (2017) 021 doi:10.1007/JHEP03(2017)021
[arXiv:1701.03480 [hep-ph]].

\bibitem{Bertone:2004pz}
G.~Bertone, D.~Hooper and J.~Silk,
Phys. Rept. \textbf{405} (2005) 279, doi:10.1016/j.physrep.2004.08.031
[arXiv:hep-ph/0404175 [hep-ph]].

\bibitem{Jungman:1995df}
G.~Jungman, M.~Kamionkowski and K.~Griest,
Phys. Rept. \textbf{267} (1996) 195, doi:10.1016/0370-1573(95)00058-5
[arXiv:hep-ph/9506380 [hep-ph]].

\bibitem{Edsjo:1997bg}
J.~Edsjo and P.~Gondolo,
Phys. Rev. D \textbf{56} (1997) 1879, doi:10.1103/PhysRevD.56.1879
[arXiv:hep-ph/9704361 [hep-ph]].

\bibitem{Baer:2018hpb}
H.~Baer, V.~Barger, J.~S.~Gainer, D.~Sengupta, H.~Serce and X.~Tata,
Phys. Rev. D \textbf{98} 075010, doi:10.1103/PhysRevD.98.075010
[arXiv:1808.04844 [hep-ph]].

\bibitem{ArkaniHamed:2006mb}
N.~Arkani-Hamed, A.~Delgado and G.~F.~Giudice,
Nucl. Phys. B \textbf{741} (2006) 108, doi:10.1016/j.nuclphysb.2006.02.010
[arXiv:hep-ph/0601041 [hep-ph]].

\bibitem{Aprile:2018dbl}
E.~Aprile \textit{et al.} [XENON],
Phys. Rev. Lett. \textbf{121} (2018) 111302, doi:10.1103/PhysRevLett.121.111302
[arXiv:1805.12562 [astro-ph.CO]].

\bibitem{Kamionkowski:1990ty}
M.~Kamionkowski and M.~S.~Turner,
Phys. Rev. D \textbf{43} (1991) 1774, doi:10.1103/PhysRevD.43.1774.

\bibitem{Gelmini:2006pw}
G.~B.~Gelmini and P.~Gondolo,
Phys. Rev. D \textbf{74} (2006) 023510, doi:10.1103/PhysRevD.74.023510
[arXiv:hep-ph/0602230 [hep-ph]].

\bibitem{Ellis:1999mm}
J.~R.~Ellis, T.~Falk, K.~A.~Olive and M.~Srednicki,
Astropart. Phys. \textbf{13} (2000) 181,
[erratum: Astropart. Phys. \textbf{15} (2001), 413-414],
doi:10.1016/S0927-6505(99)00104-8 [arXiv:hep-ph/9905481 [hep-ph]].

\bibitem{Profumo:2017ntc}
S.~Profumo, T.~Stefaniak and L.~Stephenson Haskins,
Phys. Rev. D \textbf{96} (2017) 055018, doi:10.1103/PhysRevD.96.055018
[arXiv:1706.08537 [hep-ph]].

\bibitem{Arbey:2019pdb}
A.~Arbey, J.~Ellis and F.~Mahmoudi,
Eur. Phys. J. C \textbf{80} (2020) 594, doi:10.1140/epjc/s10052-020-8152-y
[arXiv:1912.01471 [hep-ph]].

\bibitem{Drees:1992am}
M.~Drees and M.~M.~Nojiri,
Phys. Rev. D \textbf{47} (1993) 376, doi:10.1103/PhysRevD.47.376
[arXiv:hep-ph/9207234 [hep-ph]].

\bibitem{Choi:2001ww}
S.~Y.~Choi, J.~Kalinowski, G.~A.~Moortgat-Pick and P.~M.~Zerwas,
Eur. Phys. J. C \textbf{22} (2001) 563, doi:10.1007/s100520100808
[arXiv:hep-ph/0108117 [hep-ph]].

\bibitem{Drees:1993bu}
M.~Drees and M.~Nojiri,
Phys. Rev. D \textbf{48} (1993) 3483, doi:10.1103/PhysRevD.48.3483
[arXiv:hep-ph/9307208 [hep-ph]].

\bibitem{Gunion:1984yn}
J.~F.~Gunion and H.~E.~Haber,
Nucl. Phys. B \textbf{272} (1986) 1 [erratum: Nucl. Phys. B \textbf{402}
(1993), 567], doi:10.1016/0550-3213(86)90340-8.

\bibitem{Shifman:1978zn}
M.~A.~Shifman, A.~I.~Vainshtein and V.~I.~Zakharov,
Phys. Lett. B \textbf{78} (1978) 443, doi:10.1016/0370-2693(78)90481-1.

\bibitem{Bringmann:2018lay}
T.~Bringmann, J.~Edsj\"o, P.~Gondolo, P.~Ullio and L.~Bergstr\"om,
JCAP \textbf{07} (2018) 033, doi:10.1088/1475-7516/2018/07/033
[arXiv:1802.03399 [hep-ph]].

\bibitem{Ellis:2018dmb}
J.~Ellis, N.~Nagata and K.~A.~Olive,
Eur. Phys. J. C \textbf{78} (2018) 569, doi:10.1140/epjc/s10052-018-6047-y
[arXiv:1805.09795 [hep-ph]].

\bibitem{cheung2013}
C.~Cheung, L.~J.~Hall, D.~Pinner and J.~T.~Ruderman,
JHEP \textbf{05} (2013) 100, doi:10.1007/JHEP05(2013)100
[arXiv:1211.4873 [hep-ph]].
     
\bibitem{Aad:2020zxo}
G.~Aad \textit{et al.} [ATLAS],
Phys. Rev. Lett. \textbf{125} (2020) 051801, doi:10.1103/PhysRevLett.125.051801
[arXiv:2002.12223 [hep-ex]].

\bibitem{Bertolini:1990if}
S.~Bertolini, F.~Borzumati, A.~Masiero and G.~Ridolfi,
Nucl. Phys. B \textbf{353} (1991) 591, doi:10.1016/0550-3213(91)90320-W.

\bibitem{Borzumati:1994kk}
F.~Borzumati, M.~Drees and M.~M.~Nojiri,
Phys. Rev. D \textbf{51} (1995) 341, doi:10.1103/PhysRevD.51.341
[arXiv:hep-ph/9406390 [hep-ph]].

\bibitem{Okumura:2003hy}
K.~i.~Okumura and L.~Roszkowski,
JHEP \textbf{10} (2003) 024, doi:10.1088/1126-6708/2003/10/024
[arXiv:hep-ph/0308102 [hep-ph]].

\bibitem{Drees:1995hj}
M.~Drees and S.~P.~Martin,
doi:10.1142/9789812830265\_0003 [arXiv:hep-ph/9504324 [hep-ph]].

\bibitem{Dreiner:2009ic}
H.~K.~Dreiner, S.~Heinemeyer, O.~Kittel, U.~Langenfeld, A.~M.~Weber and G.~Weiglein,
Eur. Phys. J. C \textbf{62} (2009) 547, doi:10.1140/epjc/s10052-009-1042-y
[arXiv:0901.3485 [hep-ph]].

\bibitem{Choi:2007ka}
K.~Choi and H.~P.~Nilles,
JHEP \textbf{04} (2007) 006, doi:10.1088/1126-6708/2007/04/006
[arXiv:hep-ph/0702146 [hep-ph]].

\bibitem{forrester:2019co}
P.~J.~Forrester and J.~Zhang,
doi:10.2140/tunis.2021.3.55
[arXiv:1905.05314 [math-ph]].


\end{thebibliography}
\end{document}